\documentclass{aa}
\usepackage{graphicx}
\usepackage{natbib}
\usepackage{txfonts}

\bibpunct{(}{)}{;}{a}{}{,} 
\def\kms{km s$^{-1}$}
\def\mkms{\mbox{km s}^{-1}}
\begin{document}

\title{Which jet launching mechanism({\bf s}) in T~Tauri stars~?}

\author{Jonathan Ferreira\inst{1}, Catherine Dougados\inst{1} \and   Sylvie Cabrit\inst{2} } 
\institute{Laboratoire d'Astrophysique de Grenoble, BP 53, 38041 Grenoble Cedex, France\\
\email{Jonathan.Ferreira@obs.ujf-grenoble.fr} 
\and  
Observatoire de Paris, LERMA, UMR 8112 du CNRS, 61 Avenue de l'Observatoire, F-75014 Paris, France}
\offprints{J. Ferreira}

\date{Received ; accepted }
\titlerunning{Which jet launching mechanism(s) for  T~Tauri stars~?}
\authorrunning{Ferreira et al.}

\abstract
{}
{We examine whether ejection phenomena from accreting
T~Tauri stars can be described by only one type of self-collimated jet model.}
{We present analytical kinematic predictions valid soon after the Alfv\'en surface for all types of steady magnetically self-confined jets.} 
{We show that extended disc winds, X-winds, and
stellar winds occupy distinct regions in the poloidal speed
vs. specific angular momentum plane. Comparisons with
current observations of T~Tauri jets yield quantitative constraints on
the range of launching radii, magnetic lever arms, and specific energy
input in disc and stellar winds. Implications on the origin of jet asymmetries and disc magnetic fields
 are outlined.}
{We argue that ejection phenomena from accreting
T~Tauri stars most likely include three dynamical components: (1) an
outer self-collimated steady disc wind carrying most of the mass-flux
 in the optical jet (when present), confining (2) a pressure-driven coronal stellar wind and (3) a hot inner flow made of blobs sporadically ejected from the magnetopause. If the stellar magnetic
moment is parallel to the disc magnetic field, then the highly
variable inner flow resembles a "Reconnection X-wind", that has been
proven to efficiently brake down an accreting and contracting young star. If the magnetic moment is anti-parallel, then larger versions of the solar coronal mass ejections are likely to occur.
The relative importance of these three components in the observed
outflows and the range of radii involved in the
disc wind are expected to vary with time, from the stage of embedded
source to the optically revealed T~Tauri star phase.}

 
\maketitle

\section{Introduction}

Actively accreting ''classical" T~Tauri stars (TTS) often display
supersonic collimated jets on scales of a few 10-100~AU in low
excitation optical forbidden lines of [O~{\sc i}], [S~{\sc ii}], and
[N~{\sc ii}] \citep{solf93,hirt94,hirt97,lava97}, with emission
properties indicative of shock-excited outflowing gas at $\simeq
10^4$~K \citep{bacc99,lava00}. These jet signatures are correlated with the
infrared excess and accretion rate of the circumstellar disc
\citep{cabr90,hart95}. It is therefore widely believed that the
accretion process is essential to the observed jets, although the
precise physical connexion remains a matter of debate.

Over the last decade, new clues to the origin of the low-excitation
jets in TTS have been provided by sub-arcsecond observations of their
collimation and kinematic properties, using HST or ground-based
adaptive optics. A first important constraint is set by the narrow
opening angles of a few degrees observed beyond $\simeq 50$ AU of the
source \citep{burr96,ray96,doug00,hart04}.  Since TTS do not possess dense
envelopes that could confine the flow, the jets must be intrinsically
collimated. To date, the only physical process capable of producing
such unidirectional supersonic flows on the required scales is
magnetohydrodynamic (MHD) self-confinement. This is achieved by the
hoop stress due to a large scale open magnetic field anchored onto a
rotating object (see reviews by \citealt{koni00,ferr02}). However, it
remains to be established whether this MHD launching occurs
predominantly from the circumstellar accretion disc, the rotating
star, or its magnetosphere (or a combination of the above).

Interestingly, while T~Tauri stars as a group have their discs
oriented randomly with respect to the local magnetic field, those with
bright optical jets tend to have their disc axes parallel to the
ambient magnetic field direction \citep{mena04}. This trend suggests
that magnetic flux through the disc is a key parameter for the
efficiency of collimated jets in TTS, as would be expected for if jets
trace predominantly a disc wind\footnote{The term "disc wind" is 
sometimes used in the literature to refer to a thermally-driven wind,
namely an uncollimated slow outflow evaporating from the disc
surface. In this paper, disc wind refers to a magnetically driven
jet launched from the disc.}. 

Recently, it has been realized that T~Tauri jet kinematics offer a
powerful way to constrain disc wind physics. \citet{garc01b} showed
that classical "cold" self-similar disc wind solutions have
excessive terminal speeds compared to forbidden line profiles in TTS
jets, and are thus excluded, while the denser and slower "warm" disc winds are favored.
\citet{ande03} further demonstrated that the launching radius of a
keplerian disc wind may be simply derived from the jet rotation and
poloidal speeds once the stellar mass is known, regardless of the
details of the MHD disc wind solution. Interpreting the transverse
velocity shifts measured by HST at the outer edges of three T~Tauri
jets as pure rotation motions, launching radii of 0.2 to 3 AU were
inferred \citep{bacc02,ande03,coff04}, suggesting that at least the
outer portions of TTS optical jets would originate from extended
regions of the disc surface.  The small jet rotation speeds again rule out "cold" disc winds and favor 
''warm'' disc wind solutions \citep{pese04}. These results raise several important
questions:
\begin{itemize}
\item Could the same {\em extended} disc winds also explain the high-velocity
component (HVC) at 200-400 \kms\ observed in forbidden lines closer to
the jet axis (e.g. \citealt{bacc00}), or is another ejection process
necessary ? For example,  \citet{ande03} proposed that,
while the rotating outer portions of the DG Tau jet (with flow speeds $\simeq$ 45 \kms) trace a disc wind launched at 3 AU,  the HVC would originate from a separate  "X-wind" launched at the disc inner edge.  Could these two types of disc winds really coexist~?

\item Is the current precision on rotation velocities in jets sufficient 
to rule out a dominant contribution from an X-wind, stellar wind,
or magnetospheric wind to the observed optical jets ?  Are there further
diagnostics that would allow to distinguish between
these various scenarii ?
\end{itemize}

The present paper is meant to address these issues. In Section 2, we
recall the main physical ingredients of MHD ejection from the three
sites that have been proposed to contribute to optical jets
(disc, star, magnetosphere), discuss their collimation properties, and
possible coexistence. In Section 3, we generalize the work of \citet{ande03}
by presenting analytical expressions for the poloidal
and rotation speeds valid for all classes of steady, self-collimated
MHD jets. We present a diagnostic diagram in the poloidal speed
vs. specific angular momentum ($v_p$,$rv_\phi$) plane, valid soon after the
Alfv\'en surface, and show that stellar winds, X-winds and extended
disc winds follow a distinct well-defined behavior. In Section 4, we
place in this theoretical diagram the current observations of poloidal
and rotation speeds in T~Tauri jets, and discuss inferred constraints
on model parameters (launching radii, magnetic lever arm, specific
heat input). Section 5 summarizes our conclusions and their
implications.

\begin{figure*}[t]
  \includegraphics[width=\textwidth]{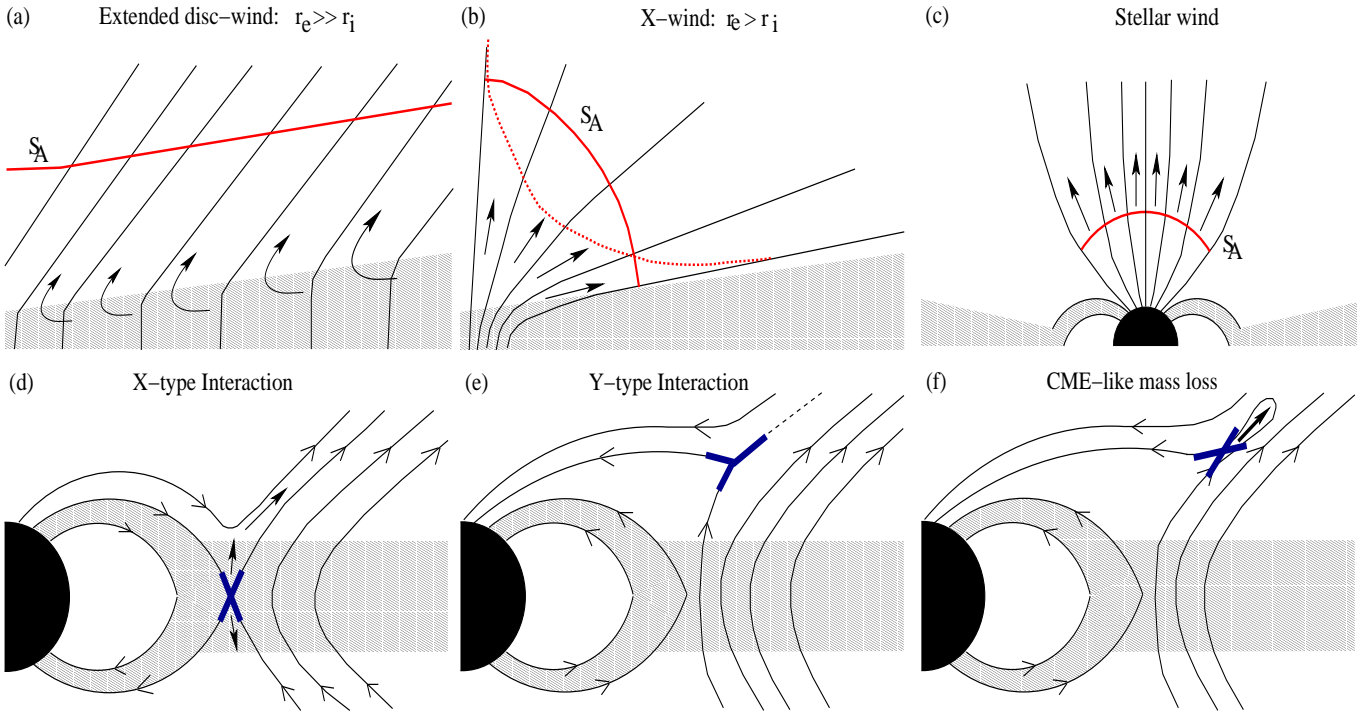}
   \caption[]{{\bf Top:} Classes of published stationary MHD jets for YSOs. When
   the magnetic field is threading the disc on a large radial
   extension (a: extended disc wind) or a small disc annulus (b:
   X-wind), jets are accretion-powered. They are mostly pressure-driven
   when the field lines are anchored onto a slowly rotating
   star (c: stellar wind). The corresponding Alfv\'en surfaces S$_A$
   have been schematically drawn (thick lines). In the X-wind case,
   two extreme shapes have been drawn: convex (solid line) and concave
   (dashed). {\bf Bottom:} Sketch of the two possible 
   axisymmetric magnetospheric configurations: (d) X-type neutral line driving 
   unsteady Reconnection X-winds, when the stellar magnetic moment is
   parallel to the disc field; (e) Y-type neutral line (akin the
   terrestrial magnetospheric current sheet) when the stellar magnetic
   moment is anti-parallel (or when the disc field is
   negligible).  (f) A CME-like ejection is produced whenever the
   magnetic shear becomes too strong in a magnetically dominated
   plasma. Such a violently relaxing event may occur with any kind of
   anti-parallel magnetospheric interaction (even with an
   inclined dipole). The thick lines mark the zones where
   reconnections occur.}
   \label{fig:mod}
\end{figure*}

\section{MHD ejection from YSOs}

Figure~\ref{fig:mod} provides a synthetic illustration of the various
configurations possibly leading to ejection in young stars.  The top
row displays steady ejection processes occurring regardless of the
interaction between the star and its circumstellar disc:
accretion-powered disc winds (Figs.~\ref{fig:mod}-a,b) and pure
stellar winds (Fig.~\ref{fig:mod}-c). The bottom row displays the
simplest possible magnetospheric star-disc interaction, namely a
dipolar magnetic field aligned parallel (Fig.~\ref{fig:mod}-d) or
anti-parallel (Fig.~\ref{fig:mod}-e,f) to the disc magnetic field. We
detail below the properties and main physical ingredients of each
ejection configuration.
 
\subsection{Accretion-powered disc winds}

In such models a large scale magnetic field is assumed to thread the
disc from its inner radius $r_i \sim r_m$ (the magnetopause) to some
external radius $r_e$ (smaller than the outer disc radius).  When $r_e
\gg r_i$ one gets a conventional "extended disc wind" as sketched in
Fig.~\ref{fig:mod}-a, for which various self-similar solutions have
been computed
\citep{blan82,ward93,ferr95,cass00a,ferr04}. Fig.~\ref{fig:mod}-b shows the
other extreme situation when $r_e \simeq r_i$, namely when ejection
occurs only from one annulus. This gives rise to a wide-angle
"X-wind" \citep{shu94a,shan02}.

Note that the only difference between these two
cases is the amount of magnetic flux threading the disc.  
Actually, they correspond to two distinct scenarii for the origin
of the large scale vertical magnetic field. In the "extended disc
wind" paradigm, the field is assumed to have been either generated by
dynamo or advected along with the accreting material (first in the
infalling stage, then in the disc stage), or both. In the X-wind
paradigm the disc is assumed to be devoid of any large scale magnetic
field.  It is further assumed that at some point in the past the
stellar magnetospheric field, which is penetrating the inner disc
regions, was flung open by outflowing disc material and that this has
led to the actual field topology where there is no causal link anymore
between the star and the disc.  Despite this difference in the origin
of the magnetic field, the ejection processes actually computed are
identical. In both situations jets carry away the exact amount of
angular momentum required to allow accretion in the underlying disc
portion\footnote{This property establishes a tight relation
between mass-flux and terminal speed, discussed
in more detail in Section~3.2.}.
However, since the range in launching radii and the shape
of the Alfv\'en surface are different, predicted terminal velocities
and angular momentum fluxes are also different and can be tested
against observations (see Section~3).

A common misconception in the literature is that "extended disc winds"
would provide only low velocities and almost no collimation. 
For example, the MHD model of \cite{ward93} matched a protostellar
disc at 100 AU to a \cite{blan82} jet model, which provided a
confinement operating only at very large scales and low terminal
speeds. However, this is not necessarily so. As shown by
\citet{cabr99} and \citet{garc01b}, full MHD solutions including the
disc wind transition \citep{ferr97} reproduce very well the observed
collimation of TTS jets and give adequate --- or even excessive ---
terminal speeds, provided their innermost launching radius $r_i$
is $\simeq$ 0.07~AU (typical disc corotation radius). 

Another issue is whether an extended disc wind and an X-wind may
physically coexist.  It has been proposed by \citet{ande03} that an
X-wind would be responsible for the HVC observed in the DG~Tau jet
whereas an extended disc wind, settled at 2-3~AU, would provide the
slower gas flowing at the outer jet edges. This proposal raises the issue of the
(vertical) magnetic field distribution within the disc. Indeed, it
would imply that, say, from the inner disc edge at $\simeq$ 0.07~AU
(locus of the X-wind) to 2-3 AU no large scale magnetic field is
present (otherwise it would drive a disc wind). It is unclear how such
a "hole" in the magnetic flux distribution could be obtained and
maintained. Besides, the collimation (hence acceleration) of the inner
X-wind would be strongly influenced by the pressure provided by the
outer disc wind, thereby modifying the results based on the current
published material.  Therefore, it appears very unlikely that an
X-wind could coexist with an extended disc wind.
However, either type of disc wind may coexist with a stellar wind,
and/or with some type of magnetospheric ejection (see below).

\subsection{MHD stellar winds}

Self-collimated stellar winds  are produced from open stellar magnetic field lines as sketched in
Fig.~\ref{fig:mod}-c. Stars rotating near break-up can provide their
rotational energy to magnetically accelerate stellar winds. But in the
case of slowly rotating young stars, such as T~Tauri stars, stellar
winds can hardly tap the protostellar rotational energy  and must be therefore essentially driven by their pressure gradient (see Section~3.3 for more details). Such a gradient stems either from thermal effects (thermally-driven winds, e.g. \citealt{saut94,saut02}) or from turbulent Alfv\'en waves (wave-driven winds, e.g. \citealt{hart80,deca81,hart82b}).   

In T~Tauri stars, accreting material is now believed to be channelled by the magnetospheric field and
to release its mechanical energy at a high latitude shock, seen mostly as
UV radiation. Under certain circumstances, some post-shocked material might keep a high temperature
and build up a large enthalpy reservoir, allowing thereby for a
thermally driven outflow. The magnetospheric accretion shock is also an efficient way to generate turbulent Alfv\'en waves. Thus, both thermal and turbulent pressures could well be present and help to launch MHD winds from the stellar surface. This would be expected to produce a correlation between accretion and ejection signatures, leading to ''accretion-powered stellar winds" (\citealt{matt05b}). 

There is indeed mounting evidence for accretion-related, hot stellar winds in T~Tauri stars 
e.g. in the form of broad blueshifted absorptions in high excitation lines \citep{edw03,dupr05},
although their relation to the low-excitation large scale jets is unclear \citep{beri01}.
We thus make the conjecture that pressure-driven stellar winds likely represent one (fast) component of observed jets, filling in their innermost part. However, we will show in Section~4.4  that such stellar winds are unlikely to carry most of the jet mass-flux, as it would require an extremely efficient conversion of thermal/wave energy into kinetic energy along the jet.

\subsection{Unsteady magnetospheric winds}

The magnetospheric interaction between the protostar and its accretion disc
provides several other possible driving mechanisms for outflows. However,
in contrast to the previous situations, these outflows are intrinsically
unsteady even in the simplified axisymmetric dipolar magnetic topology
considered here.

\subsubsection{Parallel configuration: "ReX-winds"}

Let us assume that a large scale vertical magnetic field is present in
the disc, allowing to drive an extended disc wind. If the
protostellar magnetic moment is parallel to the disc field, then a
magnetic neutral line (i.e. a true magnetic X-point in the meridional
plane) forms at the disc midplane (Fig.~\ref{fig:mod}-d) and
"Reconnection X-winds" (hereafter "ReX-winds") can be produced
above this reconnection site \citep{ferr00}. Accreted mass is lifted
vertically above the neutral line by the strong Lorentz force and is
loaded onto newly opened field lines. By this process, open field
lines carried in by the accretion flow reconnect with closed stellar
field lines. Therefore, ejected disc material is loaded onto field
lines that are now anchored to the rotating protostar. Ejection occurs
whenever the star rotates faster than the disc material, namely when
the magnetic neutral line is located farther than the corotation
radius.  Both energy and angular momentum are thereby extracted from
the rotating star and carried away in this
outflow. \citet{ferr00} showed that such a configuration provides a
very efficient spin down of a contracting protostar, much more so than any
of the other MHD ejection processes considered here.

Because of the intermittent nature of the mass loading process,
ReX-winds are best seen as a series of bullets than a
laminar flow. These bullets flow along (and push against) the inner
magnetic surface of the disc wind. This is a very interesting
situation for two reasons. First, the outer disc wind can provide also
some confinement to this inner ReX-wind. Second, the presence of a high
variability will probably lead to dissipation via shocks
(thus heating) and provide an inner pressure (against the outer disc
wind). Numerical simulations are needed to investigate such a
"two-flow" configuration (see however \citealt{hiro97,mill97}).
Note that the ReX-wind, although ejected from nearly the same
region as the X-wind (corotation), has very different characteristics,
namely:  (i) an intrinsically unsteady character,  (ii) a strong braking effect on the accreting star, (iii) the need for a large magnetic flux in the disc.

\subsubsection{Anti-parallel configuration: "CME-like ejecta"}

If, on the contrary, the stellar magnetic moment is anti-parallel to
the disc field then a magnetopause is formed with no magnetic X-point
within the disc. This situation holds in particular
in the X-wind scenario, where most of the field at the inner disc edge
is of stellar origin. The interface between the two fields adopts a Y shape
above (and below) the disc, with a neutral line all the way at high
altitudes (Fig.~\ref{fig:mod}-e)\footnote{We do not consider the case
with no large scale vertical magnetic field in the disc. Indeed, in
our opinion, it is unlikely that the central star has built up its own
magnetic field from that of the parent molecular cloud without any
field left in the circumstellar disc. However, configurations (e) and (f)
shown in Fig.~\ref{fig:mod} would also be obtained from a pure stellar
field interacting with the inner disc (see e.g. Fig.~3 in
\citealt{ostr95}).}. Such an interface does not allow to drive steady
self-collimated jets. But the presence of this neutral line is a
formidable site for time dependent energetic events, i.e
reconnections, as illustrated in Fig.~\ref{fig:mod}-f. Indeed, any
loop of stellar magnetic field that threads the disc will be sheared
by the differential rotation between the disc and the star. This
increase of magnetic energy relaxes by provoking an inflation of the
loop. If the differential rotation continues then the final stage is a
violent reconnection which leads to the ejection of a plasmoid at
roughly $45^o$ \citep{haya96,good97,matt02}. This process is somewhat
related to coronal mass ejections (CMEs) from the Sun and is believed
to explain the release of giant X-ray flares in YSOs \citep{gros04}.

One word of caution however. This physical process has only been
analytically proven in the framework of force-free fields, namely when
plasma inertia has no dynamical effect \citep{aly91}. As observed by
\citet{roma02}, when plasma inertia is important (full MHD equations),
this violent opening of the field lines is not obtained. Instead, the
magnetic configuration relaxes by modifying the rotation of the disc
thereby lowering the differential rotation. According to these
authors, all previous numerical situations showing CME-like ejecta
(eg. \citealt{matt02}) were in this magnetically dominated limit. If
this is true, then the mass flux carried by such outflows becomes an
issue if they were to explain all of the mass-loss observed in T~Tauri
jets.  In that respect, it is noteworthy that the mass-flux
ejected through this process, as modelled by \citet{matt02}, does
not seem to be correlated with the disc density, unlike what
would be needed to reproduce the accretion-ejection correlation in
T~Tauri stars.

Another issue is collimation:  $45$\degr\ is comparable to the opening 
angles of T~Tauri jets very near their base \citep{hart04}, but much larger than the
opening angles of a few degrees measured beyond
$30-50$ AU of their source \citep{burr96,ray96,doug00}. Intrinsic
collimation of CME-like ejecta cannot be invoked: Each ejected plasmoid is
made of plasma carrying its own electric currents. But because of
Ohmic resistivity (ion-electron collisions), these currents fade away
and one gets eventually unmagnetized warm gas ejected into the
interstellar medium. Without a proper global electric circuit there is
no possibility for these plasmoids to develop a self-confining
hoop-stress. Therefore, if CME-like ejecta were to explain HST jets,
they must be confined by an outer pressure operating on small scales
($< 30$ AU). A rough estimate shows that such a confinement cannot be
provided by the thermal pressure of the interstellar medium: the
medium is too cold and the density should be about $10^3$ times larger
than that of the outflow.

Alternatively, CME-like ejecta could be confined by an external magnetic
pressure. In this case $\rho_j v_j^2 \sim B_{ext}^2/\mu_o$ must hold at the
axial distance $r_j$ where confinement occurs. Outflowing mass conservation
can be written as $\dot M_j = S_j \rho_j v_j$ where $\rho_j$ is the density
of mass ejected with velocity $v_j$ through a surface of area $S_j$. Since
CME-like outflows would be produced inside a quite narrow solid angle, one
expects $S_j << 2 \pi r_j^2$. Gathering this together, one gets the
following estimate of the required magnetic field at $r_j$,
\begin{eqnarray}
B_{ext} \simeq & 25 &  \left ( \frac{\dot M_j}{10^{-9} M_{\sun}/yr} \right )^{1/2} 
 \left ( \frac{v_j}{300 \mbox{km s}^{-1}}  \right )^{1/2} \nonumber \\
 & & \times \left ( \frac{r_j}{30\mbox{ AU}}  \right )^{-1} 
 \left ( \frac{2 \pi r_j^2}{10 S_j}  \right )^{1/2} \mbox{  mG}
\end{eqnarray}
This is far above the interstellar magnetic field value and shows that
the poloidal magnetic field required to confine CME-like ejecta must
have been amplified. This can only be done if it has been advected
along by the infalling material. It must therefore be anchored on the
underlying keplerian accretion disc.  The possible presence of such a
large scale vertical magnetic field as well as its influence on the disc
dynamics (eg. launching a disc wind) have to be addressed.

To summarize, CME-like ejection raises critical issues on the mass
loss and collimation and it remains to be proven that it could
account, alone, for observed jets from T~Tauri stars. ReX-winds
have more promising properties, but they are necessarily
occurring concurrently with extended disc winds. Thus, although
time-dependent magnetospheric ejection episodes (related to reconnection
events) are most likely present in TTS, they may not dominate the
observed jet flow.

We also note that an interesting variability process could occur if
young stars sustain dynamos with global polarity inversion. In the
presence of a large scale magnetic field threading the disc, one would
indeed observe cycles of an X-type interaction with efficient ReX-wind ejection and magnetic braking, alternating with a quiescent Y-type interaction and sporadic CME-like ejections. Timescales of 10-20
yrs \citep{lope03} between jet knots might reflect such magnetic cycles.

\section{Kinematics of stationary MHD jets}

On small scales ($<$ 200 AU), spatially resolved jets from T~Tauri
stars observed in forbidden lines may be considered as
essentially steady. Indeed, even though knots are commonly observed,
the derived time scales are longer than the dynamical time scales involved in the acceleration zone. Besides, inferred shock velocities are much smaller than jet velocities
($\Delta v/v \le 25 \%$, \citealt{lava00}). Thus, steady-state models
should catch their main features. For the reasons outlined in the
previous section, we assume in the following that the mass loss in
CME-like ejecta is negligible with respect to the steady flow
components from the disc (extended or X-wind) and/or the star. The
remaining question is then: which of the latter
components dominates the jet mass flux~?

To search for possible kinematic diagnostics distinguishing among
X-wind, extended disc wind, and stellar wind, we present in this
section analytical relations for the poloidal and toroidal velocities
valid for all types of super-Alfv\'enic, stationary,
axisymmetric, self-collimated MHD jets. This is an extension of
the work of \cite{ande03}. Since all available models are
governed by the same set of ideal MHD equations, their differences
arise only from different boundary conditions. 
 
\subsection{Governing dynamical equations} 

The poloidal magnetic field writes   ${\bf B_p} = (\nabla a\times {\bf
  e_{\phi}})/r$, where $a(r,z)=cst$ describes a surface of constant
magnetic flux. An MHD jet is made of nested magnetic surfaces with several
integrals of motion along each surface. Using the usual definitions ($v_p$
poloidal velocity, $\Omega$ angular velocity and $\rho$ density), these
are: (1) the mass to magnetic flux ratio $\eta(a)$ with ${\bf v_p} = \eta
(a){\bf B_p}/\mu_o\rho$; (2) the angular velocity of a magnetic surface
$\Omega_*(a)= \Omega - \eta B_{\phi}/\mu_o\rho r$ and (3) the specific
total angular momentum $L(a)=\Omega_* r^2_A = \Omega r^2 -  rB_{\phi}/\eta$
transported away. Here, $r_A$ is the Alfv\'en cylindrical radius where mass
reaches the Alfv\'en poloidal velocity. 
The angular velocity of a magnetic surface $\Omega_*$ is roughly equal to
the angular velocity $\Omega_o$ of the mass at the anchoring cylindrical
radius $r_o$. If the magnetic field is threading the disc one gets
$\Omega_o \simeq \sqrt{GM/r_o^3}$ where $M$ is the stellar mass, whereas
$\Omega_o= \Omega_{star}$ if it is anchored into a star with a rotating
period $T =2\pi/\Omega_{star}$. The following dimensionless parameter 
\begin{equation}
\lambda = \frac{L}{\Omega_o r_o^2} \simeq \frac{r_A^2}{r_o^2} 
\end{equation} 
is a measure of the magnetic lever arm braking the rotating object (star or
disc). This parameter is an essential feature of MHD jet models, and this is
the reason why efforts are made to constrain $\lambda$ through
observations.   

We are interested here in both accretion-powered jets and stellar winds so
we need to be able to describe flows that can be substantially
pressure-driven. Thus, we allow for the presence of a heat flux ${\bf q} = \nabla H -
\nabla P/\rho$, where $H$ is the usual enthalpy for a perfect
gas. The specific heat input along any given magnetic surface is then ${\cal
  F}(s,a)=\int^s_{s_o} {\bf q} \cdot {\bf e_{\parallel}}ds'$ 
\citep{ferr04}. This term  depends on the curvilinear coordinate $s$ and therefore varies along the flow ($s_o$ is the location where mass has been loaded and ${\bf B_p} = B_p {\bf  e_{\parallel}}$).
The case of cold stellar winds driven by MHD wave pressure gradients may be treated in a similar way, by adding to the heat input ${\cal F}(s,a)$ an extra term (see e.g. Eq.~(38) in \citealt{deca81}), describing the transfer of energy from the turbulent Alfv\'en waves to the flow. We will denote ${\cal F'}(s,a)$ this generalized "pressure" term. Including this additional effect, one gets the generalized Bernoulli invariant  
\begin{equation}  
E(a) + {\cal F'}(s,a)  = \frac{v^2}{2} + H + \Phi_G - r \Omega_o
B_{\phi}/\eta 
\label{eq:bern}
\end{equation}
where $\Phi_G$ is the stellar gravitational potential and $E(a)$ is the total specific energy provided at the base of the jet  (equal to the rhs of Eq.~(\ref{eq:bern}) evaluated at $s_o$). 
Finally, the shape of each magnetic surface (therefore the whole jet) is given by the Grad-Shafranov equation, but it is of no use for obtaining the jet kinematic properties. 

Once far away from the driving source its gravitational attraction can be
neglected and the following general expressions apply 
\begin{eqnarray}
r v_\phi &=& \sqrt{GM R_o} \ \delta_o \sin^2 \theta_o \lambda_\phi
\label{eq:vphi} \\ 
v_p &=& \sqrt{\frac{G M}{R_o}} \sqrt{ \delta_o^2 \sin^2 \theta_o (2
  \lambda_p - 1) -2 + \beta } \label{eq:vp} 
\end{eqnarray}
where $R_o$ is the spherical radius ($r_o=R_o \sin \theta_o$), $\delta^2_o=
\Omega_o^2 R_o^3/GM$ is the measure of the rotation of the magnetic surface
with respect to gravity, $\beta= 2( {\cal F'} + H_o - H)/(GM/R_o)$ varies along the flow and
encompasses all pressure effects (both thermal and turbulent Alfv\'en waves) and where
\begin{eqnarray}
\lambda_\phi &= & (1 - g) \frac{r^2}{r_o^2} \nonumber \\
\lambda_p & = & \lambda_\phi \frac{1+g}{2} \\
g &=& 1 - \frac{\Omega}{\Omega_o}= \frac{m^2}{m^2-1}\left ( 1 -
\frac{r_A^2}{r^2}\right) \nonumber  
\end{eqnarray}
are signatures of magnetic effects. Here, $g$ measures the drift between
the angular velocity of the matter $\Omega$ and that of the magnetic field
$\Omega_* \simeq \Omega_o$, and $m= v_p/V_{Ap}$ is the poloidal Alfv\'enic
Mach number. These expressions are valid for any stationary MHD model of an
axisymmetric, non-relativistic jet. 

The function $g$ is a measure of the conversion of magnetic energy
into kinetic energy.  It increases from almost zero at the footpoint
to a maximum value $g_\infty$. Powerful MHD jets are those where
$g_\infty \simeq 1$ (maximum value allowed), meaning an almost
complete transfer of magnetic energy and $\lambda_p \simeq
\lambda_\phi$. In addition, these jets reach high Alfv\'enic Mach
numbers $m \gg 1$ and large radii $r \gg r_A$ so that eventually
$\lambda_\phi \simeq r_A^2/r_o^2 = \lambda $. The latter equality
means that all angular momentum has been transferred to the matter,
and is what we define as the "asymptotic regime".

At the distance from the source investigated by current observations
($\ge 30$ AU), we are likely to be in this asymptotic regime for jets
originating close to the stellar surface (stellar winds and X-winds),
while $\beta$ has become constant. In the case of disc winds
originating from a large radial extension, the asymptotic regime
is achieved further out for the outer streamlines. For instance, a
field line anchored at 1 AU has not yet reached this asymptotic regime
at $z \sim 50$~AU (see below). However, solutions that propagate far away, as demanded by observations, have $g>1/2$ already satisfied at the Alfv\'en surface (see
e.g. \citealt{ferr97}).  This is enough to obtain $\lambda_p$ and
$\lambda_\phi$ converging towards a common value, since $(1+g)/2
\rightarrow 1$ rapidly. This common value is smaller than the real
magnetic lever arm $\lambda$ (until the asymptotic regime is reached),
but it can be observationally determined (see below). We will
hereafter use $\lambda_{\phi}$ to refer to this value.

\begin{figure*}[t]
\begin{center}
\includegraphics[height=0.8\textwidth,angle=-90]{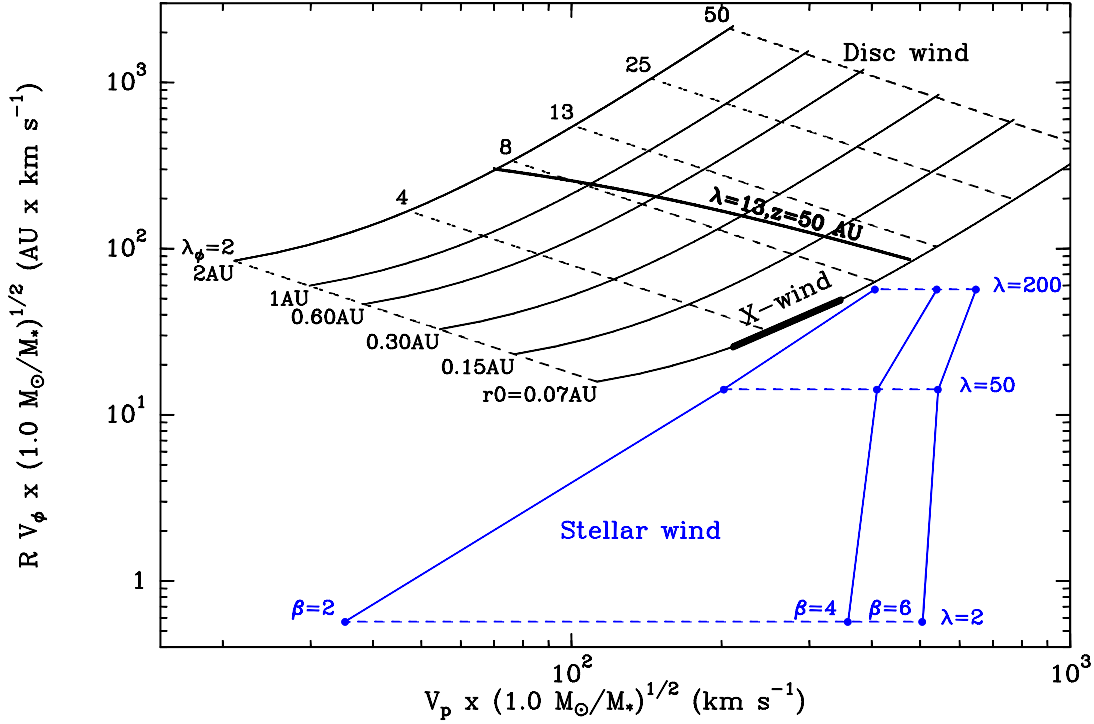}
\end{center}
   \caption{Relations between specific angular momentum vs.  poloidal
   velocity for all types of stationary MHD jet models, valid once
   $v_p \gg v_{\phi}$ (e.g. beyond the Alfv\'en surface). Plotted in
   solid lines is the relation between $rv_{\phi}$ and $v_p$ for fixed
   launching radius $r_o$ (accretion-powered disc winds;
   Eq.~\ref{eq:ander}) or fixed pressure parameter $\beta$ (stellar
   winds; Eq.~5 with typical T~Tauri parameters). Dashed lines indicate curves of constant $\lambda_{\phi} = rv_\phi/(\Omega_o r_o^2)$ in the disc wind case
   (Eq.~\ref{eq:lambda}), constant $\lambda = r_A^2/r_o^2$ in the
   stellar wind case (Eq.~4). A thick segment at $r_o = 0.07$~AU illustrates
   the locus of the X-wind (\citealt{shan98}).  We also show as a
   thick curve a cut at $z=50$ AU of a self-similar disc wind solution
   with $\lambda=13$ settled from 0.07 to 2 AU. It can be readily seen
   that $\lambda_{\phi}$ gives only a lower limit to the true
   $\lambda$ at $z=50$ AU.}
   \label{fig:rvphi_vp}
\end{figure*}

\subsection{Accretion-powered disc winds}

In the case of extended disc winds and X-winds $\delta_o = 1$ and $r_o=
R_o$ ($\theta_o = \pi/2$). We thus obtain the following expressions 
beyond the Alfv\'en surface, where $\lambda_p \simeq \lambda_\phi$:
\begin{eqnarray}
r v_\phi & = & \Omega_o r_o^2 \lambda_{\phi} \label{eq:vphi2}\\
v_p &= &\Omega_o r_o \sqrt{2 \lambda_{\phi} -3 + \beta } \label{eq:vp2}
\end{eqnarray}
The influence of pressure (thermal or waves) gradients $\beta$ in the poloidal velocity of
jets from keplerian accretion discs is barely measurable and can be
neglected. Indeed, a keplerian rotation law requires a negligible
radial pressure gradient inside the disc. As a consequence, enthalpy
at the disc surface always verifies $H_o/\Omega_o^2 r_o^2 \sim (h/r)^2
\ll 1$. Even models of highly warmed up jets, which imply huge
sub-Alfv\'enic temperatures of up to several $10^5$ K, were obtained
with $\beta$ of a few percents only \citep{ferr04}. In the case of
X-winds, the models published so far also have a negligible heat input
(see e.g. \citealt{naji94}). Thus, we can safely set $\beta = 0$ for
all accretion-powered disc winds.
 
We recall that, although thermal effects have no measurable impact on
the jet poloidal speeds, they have a tremendous one on mass loading and therefore on the kinematics. 
An important relation derived from full solutions 
of the MHD accretion-ejection flow is
\begin{equation}
\lambda \simeq 1 + \frac{1}{2\xi} \mbox{      where } \dot M_a \propto r^\xi 
\end{equation}
which relates the magnetic lever arm $\lambda$ to the ejected mass
through the local ejection efficiency $\xi$ \citep{ferr97}. The
physics are quite simple. The same torque will be applied on the
underlying accretion disc either with a large mass loss $\xi$ but
small lever arm or a small mass loss and large lever arm. What
determines exactly the maximum mass loss is the constraint to obtain
super-Alfv\'enic jets, whereas the minimum mass loss is imposed by the
disc vertical equilibrium. In vertically isothermal or adiabatic discs
("cold" models), only a tiny mass flux can be lifted from a keplerian
accretion disc, with typically $\xi \sim 0.01$ or less, giving rise to
$\lambda \sim 50$ or more \citep{ferr97,cass00a}. If there is some
heat deposition at the resistive disc upper surface layers ("warm"
models), more mass can be loaded onto the field lines, up to $\xi \sim
0.1$ or more \citep{cass00b}. These dense solutions have much smaller
magnetic lever arms $\lambda \simeq 6 - 20$.
 
Combining Eq.~\ref{eq:vphi2} and \ref{eq:vp2} with $\beta =0$ we get
\begin{eqnarray}
\frac{r v_\phi v_p}{G M} &=& \lambda_{\phi} \sqrt{2\lambda_{\phi}-3} 
\label{eq:lambda} \\
2r v_\phi \Omega_o &= & v_p^2 + 3 \Omega_o^2 r_o^2  \label{eq:ander} 
\end{eqnarray}
Eq.~\ref{eq:lambda} allows to observationally derive the value of
$\lambda_{\phi}$ independently of $r_o$ from observed values of $r
v_\phi$ and $v_p$ (assuming the stellar mass is known).  Conversely,
as first noted by \citet{ande03}, Eq.~\ref{eq:ander} is independent of
$\lambda_{\phi}$ and allows to derive the launching radius $r_o$.

It is interesting to note that \citet{ande03} derived the
Eq.~\ref{eq:ander} under the assumption that $v_\phi \ll
v_p$, while we only assumed $\lambda_p\simeq \lambda_\phi$ (i.e.
$g> 1/2$). Therefore the two assumptions must be equivalent. Since
$v_\phi \ll v_p$ is indeed observationally verified in T~Tauri
jets, we conclude that {\em (i)} Eq.~\ref{eq:ander} applies and {\em
(ii)} the plasma probed by observations has $g >$ 1/2.  
Note that, at the Alfv\'en point along any magnetic surface $v_\phi/v_p \simeq 1 - g_A$  where $g_A$ is the value of the function $g$ at this point. Thus, the plasma is also likely super-Alfv\'enic. Moreover, $g_A \simeq I_A/I_{SM}$, with the poloidal current $I= 2 \pi r B_\phi/\mu_o$.
Namely, $g$ evaluated at the Alfv\'en point measures the electrical current remaining inside the MHD
jet. Models with $g_A \sim 1$ (most of the MHD power is still available) are possible either for tiny mass fluxes or for warmed up jets \citep{cass00b}. This is a constraint for accretion-powered jet
models.

The diagram in Figure~\ref{fig:rvphi_vp} displays the curves defined
by Equations\ref{eq:lambda} and \ref{eq:ander} in the ($v_p, rv_\phi$)
plane, for $\lambda_\phi$ varying from 2 to 50 (dashed curves), and an
anchoring radius $r_o$ varying from 0.07 to 2~AU (solid curves),
illustrating the range expected for extended atomic disc
winds\footnote{The $\lambda_\phi$ values span a range corresponding to
the self-similar "warm" disc wind solutions investigated by
\citet{cass00b}. Warm solutions have been found numerically for
$\lambda \ge$ 6 but, since the condition of a positive Bernoulli
invariant requires only $\lambda > 3/2$, and $\lambda_\phi \le
\lambda$, we plot curves from $\lambda_\phi =2$ for completeness.}.

Note that as the plasma accelerates along one particular magnetic
surface, its position in the diagram moves to the right along the
corresponding $r_o=cst$ curve (once $v_p \gg v_\phi$), and
converges asymptotically towards $\lambda_\phi = \lambda$. However, an
observation is actually a cut across the jet at some altitude $z$,
i.e. it samples the whole range of nested magnetic surfaces in the
disc wind (if angular resolution is sufficient).

To illustrate the typical locus of an extended disc wind in this
diagram, we plot in Figure~\ref{fig:rvphi_vp} a transverse cut
at $z=50$ AU through a self-similar solution with $\lambda=13$, made of
surfaces anchored from $r_o=0.07$ to 2 AU (thick solid curve).
Because of self-similarity, outer streamlines reach the asymptotic
regime at larger distances than inner streamlines. As a consequence,
the discrepancy between $\lambda_\phi$ and $\lambda$ at a given
altitude is larger for outer streamlines. It reaches a factor 2 for
$r_o \ge 2$ AU at $z=50$ AU for the solution considered here. This
introduces a bias in the determination of $\lambda$, and the locus of
a self-similar disc wind deviates slightly from the theoretical
$\lambda_\phi = \lambda$ curve.

The locus of the X-wind is shown as a thick segment in Fig.~\ref{fig:rvphi_vp}. This locus is a small
fraction of that occupied by extended disc winds. As an X-wind is an accretion-powered jet launched from a tiny interval of radii $r_o$ near the inner disc
edge, the angular momentum is at the low end of the accessible range.
The range in poloidal velocities is also narrower ($\lambda \simeq
3-6$, e.g. \citealt{shan98}). There is another interesting characteristic
difference: whatever the shape of the Alfv\'en surface for the X-wind, the curve drawn by measuring $rv_\phi$ and $v_p$ at several transverse radii $r$ across the jet (at a given
altitude $z$) will always fall on the same $r_o = Cst$ curve, and will
thus have a {\em positive} slope in the ($v_p$,$rv_\phi$) plane.  
Note that the same will be true of the ReX-wind component, since it is
launched also from near corotation. On
the contrary, an extended disc wind is likely to have a {\em negative}
slope in this plane, as in the self-similar situation with constant
$\lambda$ illustrated in our $z=50$ AU cut. In order to obtain a
positive slope, the extended disc wind would need to have a markedly
larger magnetic lever arm $\lambda$ at smaller launching radii. This
is very unlikely for it would require an ejection efficiency
decreasing towards the central star (one usually expects the
contrary).

Therefore, the global amount of angular momentum, and the slope
of the $rv_\phi$ vs. $v_p$ relation across the jet at a fixed altitude
$z$, both offer good discriminants between MHD winds from the
corotation (X-wind and ReX-wind) and extended disc winds.

\begin{figure*}[t]
\begin{center}
\includegraphics[height=0.8\textwidth,angle=-90]{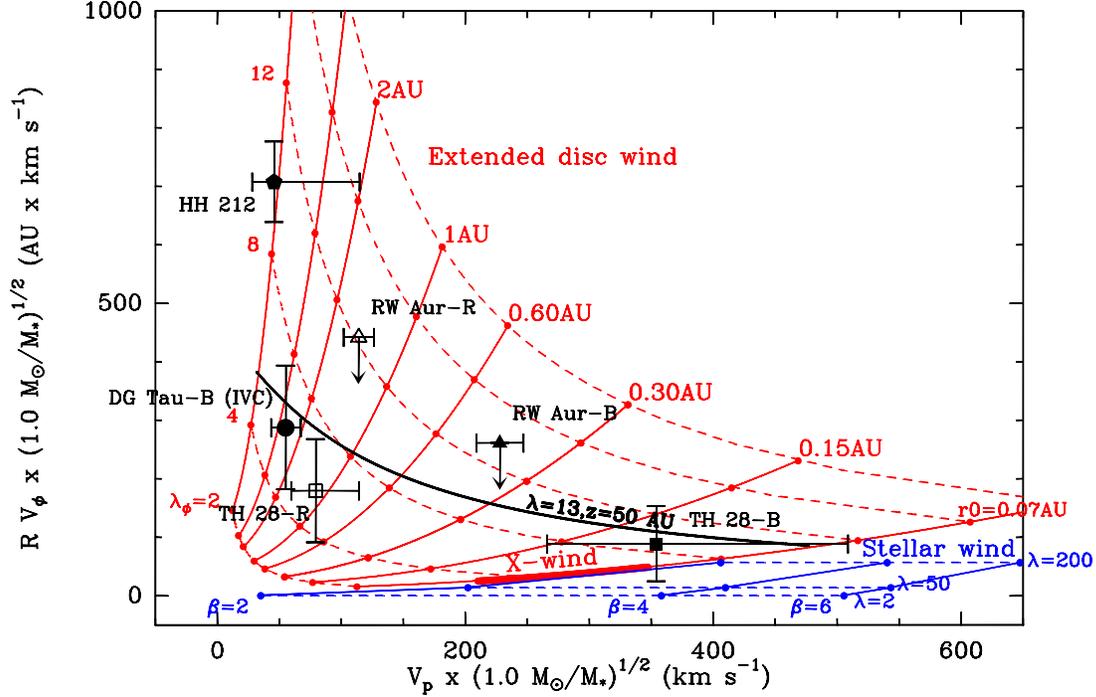}
\end{center}
   \caption{Comparison of predicted specific angular momentum vs.
   poloidal velocities with observations of T~Tauri microjets. Full
   and dashed curves show expected theoretical relations for MHD disc
   and stellar winds (same as Fig.~\ref{fig:rvphi_vp}, on a linear
   scale).  Plotted in symbols are jet kinematics measured at distance
   $z \simeq$ 50~AU in the DG~Tau, RW~Aur, and Th~28 jets. The
   infrared HH~212 jet is also shown for comparison. See text for more
   details on how the data points and their associated error bars are
   computed.}  \label{fig:rvphi_vp_obs}
\end{figure*}

\subsection{Stellar winds}

The typical rotation period of T~Tauri stars is 8 days (see \citealt{bouv97} and references therein) and corresponds to $\delta_o \simeq 0.1$. Moreover, stellar winds
are probably only launched at large latitudes ($\sin \theta_o < 1/2$),
higher anyway than those where accreting material is flowing in. Thus, Eq.~5 shows that magneto-centrifugal acceleration would be efficient only if the stellar field lines are wide open with a very large magnetic lever arm parameter, namely $\lambda > 1/\delta_o^2 \sin ^2 \theta_o \sim 200$. Although it has been recently proposed that such winds could solve the angular momentum problem in T~Tauri stars \citep{matt05b}, there is no MHD calculation yet showing the feasibility of such a flow for typical magnetic field strengths and jet mass-loss rates in T~Tauri stars. 

Instead, all self-similar models of MHD stellar winds computed so far have an Alfv\'en surface located not very far from the stellar surface. Such winds have almost straight field lines in the sub-Alfv\'enic zone so that $\lambda = r_A^2/r_o^2 \simeq R_A^2/R_o^2$, where $R_o$ is the stellar radius. For instance, model I of \citet{saut94} displays $R_A= 1.55 R_o$ ($\lambda \simeq 2.4$), whereas \citet{trus97}
obtained a solution with $R_A= 7.1 R_o$ ($\lambda \simeq 50$). As a consequence, published MHD stellar wind models are essentially pressure-driven winds where the ejected plasma gets most of its energy from $\beta$. Note that magnetic fields do play a role in collimating the outflow, but this collimation takes place mainly beyond the Alfv\'en surface. 

Figure~\ref{fig:rvphi_vp} illustrates the locus of MHD stellar winds from T~Tauri stars, using the conservative value $\delta_o \sin^2 \theta_o = 0.04$ and a stellar radius $R_o =3 R_\odot$, for $\beta$ values varying from 2 to 6 (solid lines) and $\lambda$ from 2 to
200 (dashed lines) for the sake of completeness. Stellar winds occupy only a narrow region of the ($v_p, rv_\phi$) plane of MHD models (note this is a log-log plot). They are  well separated from MHD disc winds, with only those stellar winds with very large $\lambda > 100$ approaching the X-wind region.

For a given $\lambda$ value, stellar winds are characterized by straight horizontal lines corresponding to the observable specific angular momentum:
\begin{equation}
rv_\phi  =  \delta_o \sin^2 \theta_o \lambda \sqrt{G M R_o}.
\label{eq:lambda-stellar}
\end{equation}
Figure~\ref{fig:rvphi_vp}  shows that, for published stellar wind models where $\lambda=2-50$, no detectable signature of rotation is expected from slowly rotating young stars. For the same range of $\lambda$, and contrary to accretion-powered disc winds, the asymptotic poloidal velocity of stellar winds almost entirely depends on the asymptotic value of $\beta (>2)$ alone (cf. Eq.~\ref{eq:vp}), namely
\begin{equation}
v_p  \simeq  250 \  \sqrt{\beta - 2} \left ( \frac{M}{M_{\sun}} \right
)^{1/2} \left ( \frac{R_o}{3 R_{\sun}} \right )^{-1/2} \mkms \label{eq:be}
\end{equation}
As an example, model I of \citet{saut94} used $\beta= 6.64$ giving a terminal speed of 543 \kms.
Figure~\ref{fig:rvphi_vp} clearly shows that only a much larger value of
$\lambda$ ($\ge 200$) introduces a deviation from the simple expression given by Eq.~\ref{eq:be}.

\section{Comparison with T~Tauri microjets observations}

We now compare our kinematic predictions for stationary self-confined
jet models with recent high-angular observations of the inner regions
of T~Tauri microjets, using the diagnostic diagram proposed in Figure~
\ref{fig:rvphi_vp}. The resulting implications for disc winds and stellar winds are
discussed below.

\subsection{Rotation signatures}

Detections of transverse velocity shifts suggestive of rotation
signatures have been recently derived from HST/STIS spectra at
distances $\simeq$ 50~AU from the central source in 4 T~Tauri
microjets: DG~Tau, Th~28, LkH$\alpha$~321, and RW~Aur
\citep{bacc02,coff04,woit05}.  Tentative rotation signatures were
previously reported by \citet{davi00} in the younger HH~212 jet at
2300~AU from the central infrared source (knot SK1), using the H$_2$
v=1-0 S(1) line. 

The rotation interpretation appears particularly convincing in the
case of the intermediate velocity component (IVC) of the DG~Tau jet,
as the rotation sense matches that of the disc \citep{testi02} and
transverse shift variations across and along the jet are in excellent
agreement with detailed predictions for a full MHD solution of an
extended disc wind \citep{pese04}.

Care must however be exercised as currently detected velocity shifts
are close to the detection limit and are thus affected by large
uncertainties.  Additional effects, such as intrinsic asymmetric
velocity structure in the flow and/or contamination by entrained
material, cannot be fully excluded. Detected velocity shifts may
therefore correspond only to upper limits to the true jet azimuthal
velocities. For example, tentative rotation signatures reported by
\citet{woit05} in the bipolar RW~Aur jet appear to be in opposite
sense to the disc rotation \citep{cabr06}, and are thus likely only upper
limits.

In Figure~\ref{fig:rvphi_vp_obs}, we plot the derived specific angular
momentum versus poloidal velocities for all of the above jets
(except LkH$\alpha$321, for which inclination is unknown), with
distinct symbols for redshifted and blueshifted lobes. In this graph,
we consider only significant measurements obtained towards outer jet streamlines
at transverse distances $d_\perp \ge 20$ AU from the jet axis, as
detailed modelling shows that more axial jet regions are heavily
affected by projection and beam dilution effects, leading to severe
underestimation of their azimuthal velocity \citep{pese04}. {\em As a
consequence, current observations are only probing the launching
radius $r_e$ of the outer streamlines of the optical jet}. 
 For the DG~Tau microjet, we average velocity shifts and radial
  velocities reported by \citet{bacc02} at $d_\perp = 30$~AU for distances
  along the jet between 40 and 60~AU.  For the RW~Aur microjet, we adopt
  the values of $v_\phi$ and $v_p$ published by \citet{woit05} (their table
  ~1). However, given the discrepancy between jet and disk rotation sense,
  we take these values as upper limits. For the Th~28 redshifted flow, we
  average velocity shifts and radial velocities observed by \citet{coff04}
  at $d_\perp = 34$~AU in the [O {\sc i}]6300~\AA~ and [S {\sc
  ii}]6716,6731~\AA~ lines. For the Th~28 blueshifted flow, we take the
  measurements at $d_\perp = 25$~AU in the [N {\sc ii}]6583~\AA~ line.

Table~\ref{table:obs} summarizes the adopted values of stellar mass,
jet inclination $i$, and $v_{p}$ and $v_{\phi}$ at transverse distance
$d_\perp$ together with their error bars. To place observed data points in Fig.~\ref{fig:rvphi_vp_obs}, we have taken 
into account a 10\%  (DG~Tau, RW~Aur) to 20\%  (Th~28, HH~212) uncertainty on
the value of the central stellar mass.

\begin{table*}[t]
\caption{Kinematics of T~Tauri jets with rotation estimates and derived disc wind parameters.}
\begin{center}
\begin{tabular}{lccccccccc}
\hline \hline 
Jet & M$_*$ & i & $v_{p}^{a,b}$ & $v_{\phi}^a$ & d$_{\perp}$ &
Refs.$^c$ & $\lambda_{\phi}$ & $r_e$ & $(\dot M_j/\dot M_a)_{max}^{d}$ \\ 
    & (M$_{\odot}$) & & (km s$^{-1}$) & (km s$^{-1}$) & (AU) & & & (AU) &  \\
 \hline 
DG~Tau-Blue IVC & 0.67 & 45$^{\circ}$ $\pm$ 3$^{\circ}$ &  45 $\pm$ 9  &
7.8$\pm$ 3.0 & 30. & 1 & 4-8 & 1.5-4.5 &  0.3 \\
RW~Aur-Blue & 1.3 & 46$^{\circ}$ $\pm$ 3$^{\circ}$ & 260 $\pm$ 17 & $<15.$ & 20. & 2,3 & $< 15$ & $<0.5$ & -- \\
RW~Aur-Red & 1.3 & 46$^{\circ}$ $\pm$ 3$^{\circ}$ & 130 $\pm$ 12 & $<17.$ & 30. & 2,3 & $<13$ & $<1.7$ &  -- \\
Th 28-Blue & 1: & 80$^{\circ}$ $\pm$ 3$^{\circ}$& 354$^{+150}_{-80}$  & 3.5$\pm$ 2.5 & 25. & 4 & 4-16 & 0.03-0.2 & 0.06 \\
Th 28-Red & 1: & 80$^{\circ}$ $\pm$ 3$^{\circ}$& 80$^{+33}_{-18}$ &5.3
$\pm$ 2.5 & 34. & 4 & 3.5-9 & 0.5-2 & 0.3\\
HH~212 (SK1) & 1: & 85$^{\circ}$ $\pm$ 3$^{\circ}$& 46$^{+68}_{-17}$ & 1.5 $\pm$ 0.25 & 470. & 5 & 7-17 & 2-12  & 0.2 \\
 \hline \\ 
\end{tabular}
\end{center}
$^a$Velocities $v_{p}$ and $v_{\phi}$ were computed from the observed
radial velocities and velocity shifts at transverse distance $\pm
d_\perp$ using the following expressions: $v_{p} = v_{rad}/\cos(i)$
and $v_{\phi} = v_{shift}/2\sin(i)$, where $i$ is the inclination of
the jet axis to the line of sight.

$^b$Our quoted uncertainties on $v_{p}$ take into account a typical
3$^{\circ}$ uncertainty on the inclination angle. This latter source
of error is particularly important in the case of Th~28 and HH~212
where a high inclination of 80$^{\circ}$-85$^{\circ}$ is inferred.

$^c$References: [1] Bacciotti et al. (2002); [2] Woitas et al.
(2005); [3] this paper; [4] Coffey et al. (2004); [5] Davis et al. (2000).

$^d$ one-sided mass ejection to accretion ratio.

\label{table:obs}
\end{table*}

\subsection{Constraints on wind launching radii and magnetic lever arms}

Several general conclusions may be drawn from the comparison of data
points with model predictions in Figure~\ref{fig:rvphi_vp_obs}. {\it
(i)} Dynamically cold disc winds (with magnetic lever arms
$\lambda~\ge 50$) are excluded (in agreement with the conclusions of
\citealt{garc01b,pese04}), since they would predict rotation rates largely in
excess of what is observed. {\it (ii)} X-winds and stellar winds
predict 10-100 times smaller angular momentum than suggested by
current tentative rotation signatures in TTS jets.  {\it (iii)} If
these signatures indeed trace pure rotation in the jet material, then
extended disc winds with $r_e$ of $\simeq$ 0.2 to 3 AU and moderate
magnetic lever arm parameters $\lambda_{\phi} \simeq$ 4-18
(corresponding to "warm" solutions) are needed.  In particular, it is
interesting to note that most of the current measurements, including
the upper limits in the RW~Aur jet, appear roughly compatible with the
extended MHD disc wind model with $\lambda = 13$ that fits \citep{pese04} the DG~Tau
jet dataset (thick curve in the figure).

The values of $\lambda_\phi$ and $r_e$ inferred from comparison with
disc wind predictions using Equations~\ref{eq:lambda} and
\ref{eq:ander} are listed in Table~\ref{table:obs}.  Although derived
only from the most reliable subset among available data (see previous
section), they remain within the error bars of previously published
estimates \citep{bacc02,ande03,coff04,woit05}.

We recall that if detected velocity shifts include other effects
than rotation, they give only upper limits to the true jet azimuthal
velocities. The derived launching radii and magnetic lever arms are
then also upper limits to the true disc wind parameters. Therefore,
X-winds and stellar winds cannot be definitely ruled out on the basis
of current rotation measurements which sample only the outer jet. Determination of the full transverse
jet rotation profile would be a crucial test, as demonstrated in
Section~3. This will require better angular resolution
($<$ 5~AU) and tracers of higher critical density \citep{pese04}.

\subsection{Poloidal velocities}

Independently of rotation measurements, the wide range of poloidal
velocities observed across T~Tauri jets and sometimes within the same jet (50-400 \kms) also raises
interesting issues for the jet launching mechanism. 

X-wind models launched from the disc corotation radius $r_{co}$
produce, by construction, only a relatively narrow range of poloidal
speeds. For the Alfv\'en surface calculated in e.g. \citet{shan98},
where $\lambda \simeq 3 - 6$, one obtains 
\begin{equation}
v_p \times \left({1 M_\odot/M_*}\right)^{0.5} \simeq 200-340 
\left({0.07\mbox{AU}/r_{co}}\right)^{0.5} \, \, \, \mkms  
\end{equation}
Although this range agrees with a good fraction of known T~Tauri jets
\citep{hart95}, it fails to explain the intermediate velocity
component (IVC) at $v_p < 100$ \kms\ which often dominates the emission
towards optical jet edges, as e.g. in the DG~Tau, Th~28-red, and
RW~Aur-red jets \citep{lava00,bacc00,coff04,woit05}.  Reproducing
these observations with an X-wind would require a substantial change
in the Alfv\'en surface geometry, with $\lambda \le 2$ for
equatorial streamlines. Alternatively, the IVC could be attributed to
entrainment of ambient gas. However, given the wide opening angle of
the X-wind, it is unclear whether this process could be efficient within
20-30~AU of the jet axis, where the IVC is observed. In either case, a
substantial modification of the current published solutions would be needed.

In contrast with X-winds, extended disc winds with
$\lambda \simeq 10$ easily reproduce the whole range of poloidal
speeds observed in T~Tauri jets. As can be seen in
Fig.~\ref{fig:rvphi_vp_obs}, material ejected from launching radii
beyond 0.5~AU naturally explains the intermediate velocity components
at $v_p < 100$ \kms\ seen towards jet edges, while material ejected
close to the disc corotation radius $\sim 0.07$~AU reaches high speeds
of up to 400-500 \kms\ sufficient to explain the fastest jets observed
in TTS. Hence, within a given jet lobe, a disc wind will naturally
provide a range in poloidal jet velocities because of the range of
keplerian speeds in the launching zone. Assuming $\lambda$ constant
throughout the extended disc wind, a factor 10 in launching radii will
produce a drop of a factor 3 in observed velocities, from jet axis to
jet edge. This is roughly the magnitude of the decline in poloidal
speed from axis to edge across the DG~Tau and Th~28-red jets
\citep{lava00,bacc00,coff04}.

Finally, in the case of stellar winds, Fig.~\ref{fig:rvphi_vp_obs} shows that the
range of jet poloidal speeds can easily be reproduced with $\beta
\simeq 2-4$. To get an insight of the energy reservoir required per particle, let us first assume that $\beta$ is mostly provided by an initial enthalpy. In this case, the initial jet temperature must be  
\begin{equation}  
T_o = 1.5\, 10^6\  \left ( \frac{\beta}{2} \right) \ \left (
\frac{M}{M_{\sun}} \right ) 
\left ( \frac{R_o}{3 R_{\sun} }\right )^{-1} \mbox{ K}
\end{equation}
Since  $\beta >2$ (Eq.~\ref{eq:be}), this implies that ejected material must have coronal temperatures. Such a high temperature could be provided either by a magnetic coronal activity
or by the accretion shock onto the stellar surface. Indeed, the
post-shock temperature of an adiabatic hydrodynamic shock is
\begin{equation}
  T = \frac{8}{9} \frac{m_p}{k_B} v_s^2 \simeq 9.3 \ 10^6 \left (
  \frac{v_s}{300 km s^{-1}}\right )^2 \mbox{ K}  
\end{equation}
for a fully ionized gas, where $v_s$ is the velocity of freely
falling material onto the TTS. If $\beta$ is constructed by such
a shock, then this could explain the accretion-ejection
correlation. Note however that the above equation
is an optimistic estimate, since infalling material would certainly be
cushioned by stellar magnetic fields, thus lowering the compression
and the resulting temperature. Moreover, these shocks are radiating
and matter that becomes ultimately loaded onto open field lines must
not cool too rapidly while diffusing towards the higher latitude open
field lines. TTS with prominent coronal winds do show reduced optical continuum emission from the accretion shock \citep{beri01}, in line with this kind of scenario. But the major problem with such a scenario is the unavoidable radiative losses from the densest parts:  if $\beta$ is in the form of initial enthalpy, then mass loss rates above $10^{-9} M_\odot/yr$ would radiate more X-rays than observed \citep{deca81}. 

To circumvent this problem, one has to assume that the driving pressure effects modeled by $\beta$ are mostly provided along the flow and not at the base. For instance, a sophisticated stellar wind model of \cite{saut94} has an initial temperature $T_o \sim 10^4$ K which then raises extremely sharply to $10^6$ K and levels off to several $10^6 - 10^7$ K thanks to the heating term ${\cal F}$ (see their Fig.~9). The computation of the X-ray emission of such a wind remains to be done but it is expected to be much smaller in this case. The same conclusion holds for cold wave-driven winds where $\beta\simeq 2-4$ is provided by the transfer of turbulent wave energy.

\subsection{Mass-loss rates}

As noted in Section~3.2, the magnetic lever arm in disc winds is
intrinsically related to the local ejection efficiency $\xi$ by $\lambda
\simeq 1 + 1/2\xi$ (Eq.~9). Furthermore, observations of $v_\phi$ and $v_p$ provide a lower limit
($\lambda_\phi$) on the magnetic lever arm $\lambda$, therefore an
upper limit on $\xi$. This information, combined with the knowledge of
the maximum outer anchoring radius $r_e$ (Table~1), gives an
upper limit on the (one-sided) mass ejection to accretion ratio that
can be sustained by an extended disc wind in each object:
\begin{equation}
  \frac{\dot M_j}{\dot M_a} \leq \frac{1}{4(\lambda_{\phi}-1)} \ln \frac{r_e}{r_i}.
\end{equation}
The last column in Table~\ref{table:obs} lists the maximum ratios
obtained for each of the observed TTS jets, assuming a constant
$\lambda$ throughout the disc and $r_i\sim 0.1$ AU (corotation radius
for a typical T~Tauri star).

In principle, these values could be checked for consistency against
the observationally determined ejection/accretion ratio in the same
objects. Conversely, a lower limit to the observed mass flux ratio could in
principle be used to derive an upper limit on $\lambda$ (see
\citealt{woit05}). However, such a comparison made on a star by star
basis is not as constraining as one would hope, as both jet mass-loss
rates and accretion rates currently suffer from large uncertainties:
Four different methods exist in the literature to derive jet mass-loss
rates from observed line fluxes and jet sizes, and they give results
that typically differ by a factor 5-10 (see \citealt{hart95},
\citealt{cabr02}). As for accretion rates, their derivation from
optical veiling involves a series of complex steps, with discrepancies
of up to a factor 10 between different authors
\citep{hart95,gull98}. Furthermore, veiling varies on much
shorter timescales (days) than those probed by current jet
observations ($\sim$~1 yr at $z =$50~AU), introducing an intrinsic
scatter in the ejection/accretion ratio for TTS jets.  Therefore, a
statistical comparison is probably more meaningful at this stage.  On
average, the observed ratio of (one-sided) jet mass-flux to disc
accretion rate in TTS lies in the range 0.01-0.1, depending on
the adopted accretion rates, with a scatter of a factor 3-10
\citep{cabr02}.  This range is in excellent agreement with the values
listed in Table~1, indicating that indeed magnetic disc wind models
accounting for T Tauri jet kinematics are also able to sustain the
observed jet mass-loss rates in these sources.

This is not so clear for stellar winds: The requirement that all the ejected mass gains a large specific energy through pressure gradients ($\beta$) puts a severe energetic constraint. Indeed, the total power transferred to the two jets via either thermal or wave pressure gradients is simply 
\begin{equation}
L_{\beta} \simeq \bar \beta \frac{G M \dot M_j}{R_*} \simeq \bar \beta \left (\frac{\dot M_j}{\dot M_a} \right ) L_{acc}
\end{equation}
where $\bar \beta$ is the average value of $\beta$ through the jet and $L_{acc} = G M \dot M_a/R_*$ the accretion luminosity onto the star. Taking $\bar \beta =3$ and a one-sided ejection to accretion mass ratio ranging from 1\% to 10\% gives a total power that must be as high as 3\% to 30\% of the accretion luminosity. But this transferred power is itself a fraction of the total power that must be available to the ejected material.  Since this power is presumably stored in some accretion-related turbulence, we can write $L_{\beta}= \eta L_{turb}$ where $\eta$ is the efficiency of energy conversion. For instance \citet{deca81} obtained an efficiency of only roughly 20\% with a prescribed radial field and assuming undamped waves. This is a conservative value considering that the ejected plasma will also loose energy though radiation. Thus, the total net power $L_{turb}$ that must be available for stellar winds must be as high as 15\% to 150\% of the accretion luminosity!  Clearly this is very uncomfortable. 

Note that this poses an energetic problem only if one insists on explaining all the jet mass loss with stellar winds. This probably implies that  stellar winds only carry a small fraction of the observed jet mass flux in T~Tauri stars.

\subsection{The origin of jet asymmetries}
         
An intriguing property of T~Tauri jets is the frequent occurence (in
more than half the cases) of a strong asymmetry of up to a factor 2 in
poloidal speed between the two jet lobes \citep{hirt94}, as may be
seen in Table~1 for RW~Aur and Th~28.  One might
think it difficult to obtain such asymmetries within MHD jet
models. We argue below that, in fact, all models may in principle account for
them, asymmetries representing only another set of constraints.

In the frame of stellar winds, such a velocity asymmetry could be
obtained with a difference of a factor 2 in the pressure parameter
$\beta$ between the two stellar poles (see
Fig.~\ref{fig:rvphi_vp_obs}). Further testing of this hypothesis
requires to elucidate the origin of the pressure gradients, which remains
the major unsolved issue in stellar wind models for TTS.

In the frame of MHD disc winds, jet asymmetries require an asymmetry
in magnetic lever arms, or launch radii, or both, between either sides
of the disc. Such a situation may occur naturally in an
asymmetric ambient medium.  Indeed, an accretion-ejection structure
can be seen as a rotating wheel (the disc) with two independent
electric circuits (the jets). The current that flows in each circuit
depends on the power developed by the wheel (which is fixed) but also
on the electric resistance which may be different on each side
\citep{ferr95,ferr02}.  Thus, whenever the ambient medium leads to an
asymmetric interaction (i.e. dissipation) with the bipolar jets, then
one should expect a different total current flowing at each disc
surface, namely different current densities (namely the toroidal
magnetic field, affecting $\lambda$) and/or different radial
extensions (affecting $r_e$).

Another but equivalent way to look at it is that one necessary
condition for jet production is the presence of a large scale vertical
magnetic field close to equipartition in the disc \citep{ferr95}. But
this is not a sufficient condition, as mass must be able to escape
from the turbulent accretion disc. Now, the conditions for this escape
can differ from one disc surface to the other, since they are strongly
affected by heat deposition at the surface layers \citep{cass00b}.
For instance, if the ambient radiation field is stronger in one side,
then heating and ionization can be more important, leading to enhanced
mass flux (smaller $\lambda$) and/or a larger jet launching
domain. Note also that heating of the disc surface layers could be
provided by dissipation of upstream waves, triggered by the
interaction between the jet and its environment. Any asymmetry of this
medium may therefore also lead to an asymmetric ejection.

\section{Summary and implications for the ejection mechanisms in young stars}

We have presented theoretical arguments demonstrating that unsteady MHD
ejections from the magnetosphere/disc interface, although probably
present in T~Tauri systems, may not contribute a dominant fraction of
the mass-flux in T~Tauri optical jets. 
 
To further quantify constraints on the driving mechanism for TTS jets,
we have extended upon the work of \cite{ande03}, and presented
predictions for toroidal and poloidal velocities in all types of stationary
self-collimated MHD jet models, namely accretion-driven disc winds and
pressure-driven stellar winds, valid as soon as $v_p \gg v_\phi$.

We have found that the location in the ($v_p, rv_\phi$) plane is a
general and powerful diagnostic to discriminate among MHD models and
derive relevant parameters. Comparison of model predictions with
recent observations of jet kinematics within $\sim 200$ AU of their
source yield several results:

\begin{itemize}

\item {\it Extended "cold" disc wind models are excluded in TTS}
(in line with the conclusion of \citealt{garc01b}) as their large
magnetic lever arm ($\lambda \ge 50$) predicts excessive jet rotation
on observed scales.

 \item {\it Published stellar wind models for TTS} predict a very
small specific angular momentum $\simeq 5 (\lambda/50)$ AU \kms, roughly 60-100 times smaller
than current observational estimates/upper limits in TTS jets.
Moreover, the jet poloidal speeds can only be reproduced with a large
specific energy deposition, on the order of $1-2 \times GM/R_*$. 
Such a large input of additional energy raises the question of the origin of this energy reservoir if stellar winds were to carry all the observed jet mass loss. On the other hand, there is mounting evidence for the presence of a hot inner wind close to the stellar surface in TTS  \citep{edw03,dupr05}. We propose that MHD stellar winds do contribute as an axial flow inside T~Tauri jets but carry only a small fraction of the total mass loss.

\item {\it Published wide-angle disc winds from the corotation ("X-winds")} predict a moderate specific angular momentum $\simeq 50$ AU \kms, roughly 10 times smaller than current observational
estimates/upper limits in TTS jets. Moreover, the range in poloidal
speeds is narrower than observed (50-400 \kms), and the frequent steep
decline in poloidal speed towards jet edges is not explained. Solving
these problems would require a strong modification in the Alfv\'en
surface and/or in the collimation of outer streamlines (to allow entrainment
of slow ambient gas within 20-30~AU of the jet axis).

\item {\it Extended "warm" disc winds with moderate lever arms
$\lambda \simeq 13$} \citep{cass00b} predict a range in
angular momentum and poloidal speeds readily compatible with current
observations of TTS jets, even when asymmetric, provided they are
launched from an inner radius $r_i \simeq 0.1$~AU (corotation) out to
an external radius $r_e$ ranging from 0.2 to several AU. Such extended
disc winds also reproduce the observed ejection to accretion ratios,
as well as the jet collimation and line profile shapes
\citep{pese04}. Accurate determination of the full transverse rotation
profile in the jet is however critically needed, as current
estimates of rotation speeds and launching radii are otherwise 
only upper limits. 

\end{itemize}

We therefore favor the extended disc wind scenario as the
currently simplest explanation for the main component in jets from
T~Tauri stars. Moreover, (1) it produces self-collimated jets able to
(2) provide a pressure confining any plasmoid ejected at the
magnetopause (either Reconnection X-winds or CME-like ejecta) and (3)
is hollow, allowing the propagation of an inner pressure-driven stellar
wind. We therefore expect all these components to be simultaneously
present but {\em most of the mass being carried by the disc wind}. We note
that although a stellar wind may be unable to carry a large fraction
of the ejected mass, it allows to carry the returning electric current
needed to confine the outer disc wind\footnote{The electric circuit
envisioned is flowing on the axis towards the star, enters the disc at
its inner edge (the magnetopause) and leaves it at the disc surface
(inside the jet).}. Within this framework, variability phenomena on
time scales smaller than or comparable to the stellar rotational
period should be interpreted as signatures of the star-disc
interaction. In particular, if the stellar magnetic moment is parallel to
the disc field, efficient braking of the contracting star is possible through a
ReX-wind \citep{ferr00}.

The hypothesis of extended MHD disc winds in T~Tauri stars has
several important implications on accretion disc physics.  To
obtain solutions with moderate lever arms ($\lambda \sim10$), a
necessary condition is heat input at the upper disc surface layers.
Such an input in the resistive MHD disc region allows more mass to be
loaded onto the field lines ($\xi \sim 0.1$, \citealt{cass00b}). The
origin of this heat deposition remains an open question. It 
cannot be due solely to illumination by stellar UV and X-ray radiation
(Garcia et al., to be submitted). Alternatively, the turbulent
processes responsible for the required magnetic diffusivity inside the
disc might also lead to a turbulent vertical heat flux leading to
dissipation at the disc surface layers. It is interesting to note that
in current MHD simulations of the magneto-rotational instability a
magnetically active "corona" is quickly established
\citep{ston96,mill00}. Although no 3D simulation has been done with
open magnetic field lines, this result is rather promising. Indeed, it might be an intrinsic property of the MHD turbulence in accretion discs, regardless of the launching of jets (see arguments developed by \citealt{kwan97} and  \citealt{glas04}).

The prime condition for the existence of accretion powered disc winds (extended or X-winds) is the presence of a vertical large scale magnetic field close to equipartition \citep{ferr95}, namely 
\begin{equation}
B_z \simeq  0.2\ \left ( \frac{M}{M_\odot}\right )^{1/4} 
\left ( \frac{\dot M_a}{10^{-7} M_\odot/yr} \right )^{1/2}  \left ( \frac{r_o}{1\mbox{ AU}}\right )^{-5/4 + \xi/2}  \mbox{ G,} 
 \label{eq:B}
\end{equation}
threading the disc on some extension (up to $r_e$). Thus, probing the external launching radius $r_e$ of jets from YSOs is an indirect way to constrain the disc magnetic field. The origin of this field is still a matter of debate. It could be either advected from the parent molecular cloud or locally generated by a dynamo (or both). The amount of the magnetic flux threading the disc is thus an issue in star formation, but it is likely to vary from one object to another.

\citet{ferr00} showed that contracting protostars initially rotating at break-up speeds can be spun down via ReX-winds, on the duration of the embedded phase, to low rotational speeds characteristic of TTS. But the natural outcome of this model is the decline in time of the disc magnetic flux. More precisely, the disc field reconnects with closed stellar field lines and, eventually, ends up as open magnetic lines anchored on the star. Thus, according to this picture, the relative importance of extended disc winds should decrease in time ($r_e \rightarrow r_i$). Ultimately, within this framework, some slowly rotating TTS should only drive stellar winds (along with some unsteady magnetospheric events). In this respect, it is particularly interesting to note that some TTS have indeed no spectroscopic evidence for jets in the form of high-velocity components in forbidden lines \citep{hart95}. They exhibit only a low-velocity component in [O~I] not clearly correlated with accretion, possibly tracing a photo-evaporating layer at the disc surface. Such an evolutionary effect needs testing on a broad sample of young stellar objects. Our proposed diagram will then be a prime diagnostic tool to use.

\begin{acknowledgements}
SC's visits to LAOG were partly funded by the Programme National de Physique Stellaire. The authors acknowledge support through the Marie Curie Research Training Network JETSET (Jet Simulations, Experiments and Theory) under contract MRTN-CT-2004-005592. 
\end{acknowledgements}

\end{document}